\documentclass[aps,pra,twocolumn,amsmath,amssymb, superscriptaddress,tightenlines]{revtex4}
\usepackage{graphicx}% Include figure files
\usepackage{epsfig}
\usepackage{dcolumn}% Align table columns on decimal point
\usepackage{bm}% bold math
\usepackage{amssymb}
\usepackage{bm}
\usepackage{amsmath, bm}
\usepackage{upgreek}
\usepackage{enumitem}  
\usepackage[colorlinks,citecolor=blue,linkcolor=blue,hyperindex]{hyperref}

\begin{document}

\title{Anyon condensation, topological quantum information scrambling, and Andreev-like reflection of non-Abelian anyons in quantum Hall interfaces}
\author{Ken K. W. Ma}
\affiliation{National High Magnetic Field Laboratory, Tallahassee, Florida 32310, USA}
\date{\today}

%----------------------- Abstract ------------------------------

\begin{abstract}

Quantum information scrambling is the spread of local information into correlation throughout the entire quantum many-body system. This concept has become a central topic in different contexts. In this work, we restate the connection between anyon condensation and topological quantum information scrambling in quantum Hall interfaces. We consider the interface between the Abelian Halperin-330 state and the non-Abelian Read-Rezayi state. We verify explicitly that the interface can be fully gapped. This allows the transmutation of local pseudospin information carried by an Abelian anyon into topological information stored entirely by the anyons in the non-Abelian quantum Hall liquid, with no scrambled information stored at the interface. In combination with our previous work~\href{https://link.aps.org/doi/10.1103/PhysRevB.105.045306}{[K. K. W. Ma and K. Yang, Phys. Rev. B \textbf{105}, 045306 (2022)]}, our results demonstrate the dependence of the scrambling mechanism on the gapfulness of the interface. Possible Andreev-like reflection of non-Abelian anyons in the fully gapped interface is also discussed. 

\end{abstract}

\maketitle

%-------------------------- Introduction -----------------------------

\section{Introduction}

Quantum information scrambling describes the dispersal of local information into entanglement and correlation throughout the entire system. As a result, the original information is stored nonlocally and cannot be recovered via local measurement. This concept was originally introduced to study black hole physics~\cite{Hayden2007, Susskind2008, Susskind2011, Hayden2013, Stanford2014, Stanford2015}. Later, it has attracted considerable attention in the context of many-body dynamics~\cite{Stanford2016, Yoshida2016, Yoshida2017, Sagawa2018, Yao2019} and quantum neural network~\cite{Zhai2020, Zhai2021, Jaffe2022}. Furthermore, the concept has been verified in various recent experiments~\cite{Landsman2019, Pan2022}.

In a seemingly different context, our recent work~\cite{Pf-331} studied quasiparticle tunneling across an interface between two different quantum Hall (QH) states at the \textit{same} Landau-level filling factor. We showed that an original Abelian anyon from the Halperin-331~\cite{Halperin} QH liquid would transmute into a pair of anyons, when it crosses the interface and enters the non-Abelian Pfaffian (or Moore-Read) QH liquid~\cite{MR1991}. One of the resulting anyons is neutral and being created at the interface, and the remaining one is created in the Pfaffian liquid. Consequently, the original pseudospin information carried by the Abelian anyon is stored nonlocally, and becomes inaccessible in any local measurement. Thus, the quasiparticle transmutation can be viewed as a kind of quantum information scrambling. It is remarked that the scrambling here is entirely topological, which does not involve any thermalization in the system.

Recent studies have revealed that the physics of QH interfaces is more complicated yet much richer than the standard QH edge physics~\cite{Pf-331, Grosfeld2009, Bais-PRL2009, wan16, Yang2017, Santos2017, Mross, Wang, Lian, simon20, zhu20, Hughes2019,  Regnault1, Regnault2, Nielsen1, Nielsen2, Teo2020, Heiblum2021, Mross2021, QH-interface2021, Oguz2022}. A very recent surprise is that the dynamics of QH interface is actually described by a nonrelativistic string-like theory~\cite{QH-interface2021, Oguz2022}. By proximitizing different QH liquids, a wide variety of interfaces can be formed. The corresponding quasiparticle transmutation (if any) will depend on the properties of the interface. Hence, different ways of scrambling the original information may be realized in different QH interfaces. The previously studied 331-Pfaffian interface has a chiral central charge of $c=2-3/2=1/2$, indicating that there must be a gapless chiral Majorana fermion mode on the interface~\cite{Yang2017}. As a result, some of the scrambled information can be stored at the interface. 

On the other hand, an interface with a zero chiral central charge can be created by proximitizing a pair of different QH states with identical central charges for their edges. Interfaces of this kind have been studied extensively, and importantly, can be gapped or gapless. When both QH phases are Abelian, they are described by the $K$-matrix formalism~\cite{Wen1992-K, Wen-book, footnote-Lan}. This enables one to study the gapfulness of the interface through the concepts of Lagrangian subgroups~\cite{Kapustin2011, Levin2013, Qi2013a, Qi2013b, Kapustin2014, Juven2015} or the null vector criteria~\cite{Haldane1995}. For an interface involving non-Abelian QH state(s), its gapfulness and the corresponding gapped phases can be explored by the anyon condensation approach~\cite{Bais-PRB2009, Fuchs2013, Ellens2014, Kong2014, Hung2014, Hung2015, Bernevig, Chenjie2017, Burnell-review}. A rigorous mathematical formalism was developed in~\cite{Wen2015, Wen2020}. Meanwhile, gapped interfaces between different non-Abelian topologically ordered states is still an ongoing research topic. Besides the mathematical quest, it is tempting to explore the related exotic physical phenomena in such a special kind of interface. 

Motivated by the above discussion, we consider in this work the interface between the Abelian Halperin-330 state~\cite{Halperin} and the non-Abelian Read-Rezayi state at level four (in short, the RR$_4$ state)~\cite{RR-state}. Both states or phases may describe the fractional QH (FQH) state in a bilayer system at the total Landau-level filling factor $\nu=2/3$. Whether these phases are favorable or not in a realistic sample depends on the actual microscopic details of the system. Here, we assume both states and the interface between them can be realized, and study the possible consequences. Previous theoretical work suggested that a continuous phase transition between the Halperin-330 state and the RR$_4$ state might be triggered by tuning the interlayer tunneling strength in a bilayer FQH system~\cite{preprint2010, Wen-PRL2010, Wen-PRB2010, Wen-PRB2011}. This transition and similar phase transitions in bilayer systems at other filling factors can be studied systematically through the anyon condensation approach~\cite{Wen-PRL2010, Wen-PRB2012}. In particular, the condensation of a specific type of anyon in the RR$_4$ state leads to the Halperin-330 state. Based on the ``folding trick"~\cite{Kong2014, Hung2015, Chenjie2017}, it is expected that a gapped boundary between these two FQH states can form. Nevertheless, the precise gapping mechanism and the possibility of having different phases for the gapped interface have not been addressed. Suppose the interface can be gapped, the scrambled pseudospin information may be entirely stored by anyons in the non-Abelian RR$_4$ FQH liquid with no information stored at the interface. This feature (or scrambling mechanism) is completely different from the one in the aforementioned 331-Pfaffian interface~\cite{Pf-331}. Furthermore, the Halperin-330 state and the RR$_4$ state host different sets of anyons. This leads to the possibility of realizing Andreev-like reflection of non-Abelian anyons at the interface, which we will also discuss. While most of the techniques and some results in this work will resemble those in Refs.~\cite{Wen-PRL2010, Wen-PRB2011}, we aim at providing a pedagogical discussion of the rich set of physical phenomena that gapped interfaces involving non-Abelian FQH states can offer. This serves as an example of the connection between the somehow abstract mathematical formulation and realistic physical effects in gapped boundaries of topologically ordered states.

\section{Review of Halperin-330 and Read-Rezayi states}
\label{sec:330-RR}

In this section, we summarize some basic results from the theory of topological order~\cite{Wen1990, Wen-book} and the conformal field theory (CFT) approach~\cite{Hansson-CFT} for studying FQH effect. This is followed by an overview of the Halperin-330 state and the Read-Rezayi state (RR state). The review here aims at introducing necessary notations and setting the stage for our later discussion.

\subsection{Basic facts for Abelian FQH states}

In general, different operations between anyons are described by the algebraic theory of anyons~\cite{Kitaev2006}. For a pair of anyons $a$ and $b$ in a topological order $\mathcal{A}$, their fusion leads to another anyon $c\in\mathcal{A}$. Mathematically, the fusion is described by 
\begin{eqnarray} \label{eq:fusion}
a\times b=\sum_c N^{c}_{ab} c.
\end{eqnarray}
An anyon $a$ is Abelian if and only if its fusion with any $b$ gives a unique outcome (i.e., no summation in $c$). If $\mathcal{A}$ consists of Abelian anyons only, then it is an Abelian topological order. Furthermore, the mutual statistics between the two anyons $a$ and $b$ (defined as exchanging the anyons twice) is captured by the monodromy,
\begin{eqnarray} \label{eq:monodromy}
M^{ab}_{c}=\exp{[2\pi i(s_c-s_a-s_b)]}.
\end{eqnarray}
This corresponds to the accumulation of a braiding phase of $2\pi(s_c-s_a-s_b)$. Here, $s_a=h_a-\bar{h}_a$ is the conformal spin of the anyon $a$, which has the conformal weights $(h_a, \bar{h}_a)$ for its associated CFT operator. Without any confusion, we always refer to the CFT operator when we use the terminologies ``conformal spin" and ``scaling dimension" of an anyon. Since the braiding phase generally depends on the fusion channel $c$, it is important to specify $c$ in the braiding of non-Abelian anyons. Moreover, the scaling dimension of the anyon $a$ is $h_a+\bar{h}_a$. In a holomorphic or chiral CFT, $\bar{h}_a=0$.  The possible types of anyons and their representations in CFT depend on the actual physical system. For a pair of anyons $a$ and $b$ with scaling dimensions $h_a$ and $h_b$, the operator product expansion (OPE) bewteen their CFT operators is~\cite{Hansson-CFT, CFT-book}
\begin{eqnarray} \label{eq:general-OPE}
\lim_{z\rightarrow w} \mathcal{O}_a(z)\times \mathcal{O}_b(w)
\sim\sum_c \left(z-w\right)^{h_c-h_a-h_b}\mathcal{O}_c (w).
\end{eqnarray}
Here, $c$ are the possible anyons resulting from fusing $a$ and $b$ [see Eq.~\eqref{eq:fusion}]. The OPE describes the singular behavior when the two fields (equivalently, the corresponding anyons) come close to each other.

Suppose the FQH state can host Abelian anyons only. Then, its effective low-energy theory is described by an Abelian topological order. Moreover, the low-energy physics of the system is dominated by its gapless edge, which can be described by the Lagrangian density~\cite{Wen-book},
\begin{eqnarray} \label{eq:edge-L}
L_{\text{edge}}
=-\frac{1}{4\pi}
\sum_{i,j}
\left(
 K_{ij} \partial_t\phi_i\partial_x\phi_j
+V_{ij} \partial_x\phi_i \partial_x\phi_j\right).
\end{eqnarray}
Here, $\phi_i$ denotes the chiral Bose modes on the edge. The index $i$ runs from 1 to $N$, where $N$ is the number of edge modes. The chiralities of these modes are determined by the signs of the eigenvalues of the $K$ matrix. Since different information of the topological order is encoded in the $K$ matrix and the associated $t$ vector~\cite{footnote-spin}, the first term in Eq.~\eqref{eq:edge-L} is known as the topological term. In particular, the filling factor of the corresponding QH liquid is given by $\nu=t^T K^{-1}t$, where $t^T$ denotes the transpose of $t$. The interaction between different edge modes is described by the term that involves the $V$ matrix. 

Different anyons in the topological order (or quasiparticles and quasiholes in the FQH system) are created by the vertex operator $\mathcal{V}_l=:\exp{(il^T\phi)}:$~\cite{CFT-book, Hansson-CFT}. Here, $\phi$ is a column vector with elements $\phi_i$ in Eq.~\eqref{eq:edge-L}. The symbol $:\mathcal{O}:$ stands for the normal ordering of $\mathcal{O}$, which will be skipped below. In order to be a legitimate anyon, it must braid trivially with all possible electrons in the FQH state. This translates into the requirement that the OPE between its associated CFT operator and all possible electron operators are singlevalued. Suppose the anyon is legitimate, then its charge is given by $Q=e(l^T K^{-1}t)$. When the edge is maximally chiral (i.e., all edge modes have the same chirality), then the operator has a universal scaling dimension $h_l=(1/2) (l^T K^{-1}t)$. When the edge has counterpropagating edge modes, the scaling dimension becomes non-universal, and depends on the interaction between the edge modes~\cite{Wen-book}.

\subsection{Overview of the Halperin-330 state}

The Halperin-330 state is a two-component Abelian topological order~\cite{Halperin}. It is characterized by the matrix,
\begin{eqnarray}
K=\begin{pmatrix}
3 & 0 \\ 0 & 3
\end{pmatrix},
\end{eqnarray}
and the associated $t=(1,1)^T$. Clearly, one has $\nu=t^T K^{-1}t=2/3$. In a more explicit form, the corresponding topological term for the Halperin-330 edge is
\begin{eqnarray} \label{eq:L0-330}
L_0
=-\frac{3}{4\pi}\left(\partial_t\phi_\uparrow\partial_x\phi_\uparrow
+\partial_t\phi_\downarrow\partial_x\phi_\downarrow\right).
\end{eqnarray}
Here, the subscripts $\uparrow, \downarrow$ denote the layer or the pseudospin index. Roughly speaking, each layer of the Halperin-330 state is a simple Laughlin state at 
$\nu=1/3$ ~\cite{Laughlin1983}. Since $\phi_\uparrow$ and $\phi_\downarrow$ have the same chirality, the edge is maximally chiral and has a central charge $c=2$. 

The two most relevant vertex operators for electrons are $\exp{(3i\phi_\uparrow)}$ and $\exp{(3i\phi_\downarrow)}$. These two different operators create respectively an electron in the upper and the lower layer. We call them as electrons with pseudospin up and pseudospin down. Both electrons operators have scaling dimensions $3/2$, indicating that they are indeed fermionic as required. The most fundamental anyon that the Halperin-330 state can host has charge $e/3$. Depending on its pseudospin, it is created by the operator $\exp{(i\phi_\uparrow)}$ or $\exp{(i\phi_\downarrow)}$. The OPE between any one of them and each of the electron operators is singlevalued. For example, one has the following OPEs:
\begin{align}
\lim_{z\rightarrow w}
e^{3i\phi_\uparrow(z)}\times e^{i\phi_\uparrow(w)}
&\sim (z-w)e^{4i\phi_\uparrow(w)},
\\
\lim_{z\rightarrow w}
e^{3i\phi_\downarrow(z)}\times e^{i\phi_\uparrow(w)}
&\sim e^{i[3\phi_\downarrow(w)+\phi_\uparrow(w)]}.
\end{align}
Notice that $|\alpha|=1$ is the smallest possible nonzero value for the generic operator $\exp{(i\alpha\phi_\uparrow)}$ to have singlevalued OPEs with both electron operators. This fact verifies that the $e/3$ anyon is the smallest-charge anyon that can be host by the Halperin-330 state.

\subsection{Overview of the Read-Rezayi states}

In contrast to the Halperin-330 state, the RR states are non-Abelian topological orders introduced from the CFT approach~\cite{RR-state}. Each RR state consists of two different types of CFTs. First, it has a compactified U(1) holomorphic boson $\phi$. On the edge of the system, $\phi$ corresponds to the gapless charge mode described by
\begin{eqnarray}
L_\phi
=-\frac{1}{4\pi\nu}\partial_x\phi(\partial_t+v\partial_x)\phi.
\end{eqnarray}
This mode fixes the charge density $\rho(x)=\partial_x\phi/2\pi$ and the quantized electrical Hall conductance $\sigma_{xy}=\nu e^2/h$. The possible values of the filling factor $\nu$ depend on the second type of CFT in the RR state, which is the chiral $\mathbb{Z}_k$ parafermion CFT with the central charge~\cite{ZF-parafermion1985, CFT-book},
\begin{eqnarray}
c=\frac{2(k-1)}{k+2}.
\end{eqnarray} 

The above parafermion CFT can be obtained from the SU(2)$_k$/U(1) coset construction~\cite{Gepner1987}. Here, $k\in\mathbb{N}$ is the level of the corresponding Wess-Zumino-Witten model. Following the notations in Ref.~\cite{Slingerland-Bais} (which are more convenient for the general discussion), each Virasoro primary field in the parafermion CFT is labeled as $\Phi^\ell_m$ with $\ell+m\equiv 0~(\text{mod } 2)$. The index $\ell=0,1,\cdots, k$. By imposing the field identification, $\Phi^\ell_m=\Phi^\ell_{m+2k}=\Phi^{k-\ell}_{m-k}$, the index $m$ can be restricted to the range $-\ell<m\leq \ell$, and $\ell>0$. Hence, there are in total $k(k+1)/2$ Virasoro primary fields. Their scaling dimensions are given by
\begin{eqnarray}
h^\ell_m=\frac{\ell(\ell+2)}{4(k+2)}-\frac{m^2}{4k}.
\end{eqnarray}
Since the CFT is chiral or holomorphic, the scaling dimension of the field is also its conformal spin. The fusion rule between different fields is
\begin{eqnarray}
\Phi^{\ell_1}_{m_1}\times\Phi^{\ell_2}_{m_2}
=\sum^{\min{(\ell_1+\ell_2, 2k-\ell_1-\ell_2)}}_{\ell=|\ell_1-\ell_2|}\Phi^\ell_{m_1+m_2}.
\end{eqnarray}
When $k\geq 2$, the $\mathbb{Z}_k$ parafermion CFT is non-Abelian. A famous example is the Ising CFT at $k=2$, which has a non-Abelian anyon $\sigma$ corresponding to the field $\Phi^1_1$. The Ising CFT is relevant in the description of FQH state at $\nu=5/2$~\cite{52-review2022}. The parafermion CFTs and their related FQH states have attracted considerable amount of attention as the non-Abelian anyons there are useful in topological quantum computation~\cite{Freedman2002, Kitaev2003, Bonesteel2005, TQC-RMP2008, Simon2007-TQC, Werner2009, Hutter2016, Pachos2012, Pachos2017}. 

An anyon (including the electron) in the RR state is created by the operator, 
$\eta\exp{(i\omega\phi)}$. Here, $\eta$ is a Virasoro primary field in the parafermion CFT. It is customary to choose $\eta=\Phi^k_{k-2}$ for constructing the electron operator. The conformal spin for $\Phi^k_{k-2}e^{i\phi/\nu}$ is
\begin{eqnarray}
h=\frac{k-1}{k}+\frac{1}{2\nu}.
\end{eqnarray}
By requiring $\nu>0$ and $h$ being an half-integer, the possible filling factor is fixed at $\nu=k/(Mk+2)$, where $M$ is a positive odd integer. For $k=4$ and $M=1$, the Read-Rezayi state may describe a FQH state of electrons at $\nu=2/3$. This matches the filling factor of the Halperin-330 state. In particular, the Lagrangian density describing the chiral Bose mode $\phi_l$ on the RR$_4$ edge is
\begin{eqnarray} \label{eq:edge-RR4}
L_{\phi_l}=-\frac{3}{8\pi}\partial_x\phi_l(-\partial_t+v\partial_x)\phi_l.
\end{eqnarray}
Since we will eventually study the interface between two FQH liquids, we define the chirality of $\phi_l$ as opposite to the chiralities of the two Bose modes on the Halperin-330 edge. The subscript $l$ indicates that $\phi_l$ is a left-moving mode. As a reminder, the complete edge theory for the RR$_4$ state also involves the neutral parafermionic sector (see the particular discussion in Ref.~\cite{Bishara2008}). 

For later reference, the ten different Virasoro primary fields in the $\mathbb{Z}_4$ CFT and their fusion rules are listed in Table~\ref{tab:Z4-fusion}. Note that we have switched to another set of notations (somehow more convenient for this particular discussion) for the primary fields. The identification with the notations in the previous discussion:
\begin{align}
\nonumber
I&=\Phi^4_4, \psi_1=\Phi^4_2, \psi_2=\Phi^4_0, \psi_3=\Phi^4_{-2}, 
\sigma_+=\Phi^3_3,
\\
\sigma_-&=\Phi^1_1, \epsilon=\Phi^2_0, \rho=\Phi^2_2, 
\chi_-=\Phi^3_1, \chi_+=\Phi^3_{-1}.
\end{align}
After determining the electron operator, all other operators for quasiparticles can be deduced in a ``brute-force" manner by requiring the OPE between $\eta \exp{(i\omega\phi)}$ and the electron operator is singlevalued. This restricts the possible values of 
$\omega$ in the vertex operator for each separate $\eta$ as listed in Table~\ref{tab:Z4}. From the result, the charge of the corresponding quasiparticles $2\omega/3$ (in unit of $e$) and the whole spectrum of anyons that the Read-Rezayi state can host is determined. Notice that the quantum dimension of the anyon $a$ is determined by the largest eigenvalue of the matrix $\bm{N}_a$ with matrix elements $(\bm{N}_a)_{bc}=N^c_{ab}$ in Eq.~\eqref{eq:fusion}. This maximum eigenvalue is positive and nondegenerate, as guaranteed by the Perron-Frobenius theorem. For an Abelian anyon, the quantum dimension is one. Otherwise, the anyon is non-Abelian.

\begin{table*} [htb]
\normalsize
\begin{center}
\begin{tabular}{|c|c|c|c|c|c|c|c|c|c|c|}
\hline
~$\times$~ & ~$I$~ & ~$\psi_1$~ & ~$\psi_2$~ & ~$\psi_3$~ & ~$\sigma_+$~ & ~$\sigma_-$~ & ~$\epsilon$~  & ~$\rho$~ & ~$\chi_+$~ & ~$\chi_-$~   
\\ \hline
~$I$~ &  ~$I$~ & ~$\psi_1$~ & ~$\psi_2$~ & ~$\psi_3$~ & ~$\sigma_+$~ & ~$\sigma_-$~ & ~$\epsilon$~  & ~$\rho$~ & ~$\chi_+$~ & ~$\chi_-$~   
\\ \hline
~$\psi_1$~ & ~$\psi_1$~ & ~$\psi_2$~ & ~$\psi_3$~ & ~$I$~ & ~$\chi_-$~ & ~$\sigma_+$~ & ~$\rho$~  & ~$\epsilon$~ & ~$\sigma_-$~ & ~$\chi_+$~   
\\ \hline
~$\psi_2$~ & ~$\psi_2$~ & $\psi_3$ & ~$I$~ & ~$\psi_1$~ & ~$\chi_+$~ & ~$\chi_-$~ & 
~$\epsilon$~ & ~$\rho$~  & ~$\sigma_+$~ & ~$\sigma_-$~   
\\ \hline
~$\psi_3$~ & ~$\psi_3$~ & $I$ & ~$\psi_1$~ & ~$\psi_2$~ & ~$\sigma_-$~ & ~$\chi_+$~ & 
~$\rho$~ & ~$\epsilon$~  & ~$\chi_-$~ & ~$\sigma_+$~   
\\ \hline
~$\sigma_+$~ & ~$\sigma_+$~ & $\chi_-$ & ~$\chi_+$~ & ~$\sigma_-$~ & 
~$\psi_1+\rho$~ & ~$I+\epsilon$~ & ~$\sigma_++\chi_+$~ & ~$\sigma_-+\chi_-$~  & 
~$\psi_3+\rho$~ & ~$\psi_2+\epsilon$~   
\\ \hline
~$\sigma_-$~ & ~$\sigma_-$~ & $\sigma_+$ & ~$\chi_-$~ & ~$\chi_+$~ & 
~$I+\epsilon$~ & ~$\psi_3+\rho$~ & ~$\sigma_-+\chi_-$~ & ~$\sigma_++\chi_+$~  & 
~$\psi_2+\epsilon$~ & ~$\psi_1+\rho$~   
\\ \hline
~$\epsilon$~  & ~$\epsilon$~ & $\rho$ & ~$\epsilon$~ & ~$\rho$~ & 
~$\sigma_++\chi_+$~ & ~$\sigma_-+\chi_-$~ & ~$I+\psi_2+\epsilon$~ & 
~$\psi_1+\psi_3+\rho$~  & ~$\sigma_++\chi_+$~ & ~$\sigma_-+\chi_-$~   
\\ \hline
~$\rho$~ & ~$\rho$~ & $\epsilon$ & ~$\rho$~ & ~$\epsilon$~ & 
~$\sigma_-+\chi_-$~ & ~$\sigma_++\chi_+$~ & ~$\psi_1+\psi_3+\rho$~ & 
~$I+\psi_2+\epsilon$~  & ~$\sigma_-+\chi_-$~ & ~$\sigma_++\chi_+$~   
\\ \hline
~$\chi_+$~ & ~$\chi_+$~ & $\sigma_-$ & ~$\sigma_+$~ & ~$\chi_-$~ & 
~$\psi_3+\rho$~ & ~$\psi_2+\epsilon$~ & ~$\sigma_++\chi_+$~ & 
~$\sigma_-+\chi_-$~  & ~$\psi_1+\rho$~ & ~$I+\epsilon$~   
\\ \hline
~$\chi_-$~  & ~$\chi_-$~ & $\chi_+$ & ~$\sigma_-$~ & ~$\sigma_+$~ & 
~$\psi_2+\epsilon$~ & ~$\psi_1+\rho$~ & ~$\sigma_-+\chi_-$~ & 
~$\sigma_++\chi_+$~  & ~$I+\epsilon$~ & ~$\psi_3+\rho$~   
\\ \hline
\end{tabular}
\caption{Fusion rules for the 10 primary fields in the $\mathbb{Z}_4$ parafermion CFT}
\label{tab:Z4-fusion}
\end{center}
\end{table*}

\begin{table*} [htb]
\begin{center}
\begin{tabular}{| l | c | c | c | c | c | c | c | c | c | c |}
\hline
~Primary field ($\eta$)~ & ~$I$~ & ~$\psi_1$~ & ~$\psi_2$~ & ~$\psi_3$~ & ~$\sigma_+$~ & ~$\sigma_-$~ & ~$\epsilon$~  & ~$\rho$~ & ~$\chi_+$~ & ~$\chi_-$~   
\\ \hline
~Conformal dimension $(h_\eta)$~ & ~$0$~ & ~$3/4$~ & ~$1$~ & ~$3/4$~ & ~$1/16$~ & ~$1/16$~ & ~$1/3$~ & ~$1/12$~ & ~$9/16$~ & ~$9/16$~ 
\\ \hline
~Quantum dimension $(d_\eta)$~ & ~$1$~ & ~$1$~ & ~$1$~ & ~$1$~ & ~$\sqrt{3}$~ & ~$\sqrt{3}$~ & ~$2$~ & ~$2$~ & ~$\sqrt{3}$~ & ~$\sqrt{3}$~ 
\\ \hline
~Possible values for $\omega_\eta$ & ~$m$~ & ~$m+1/2$~ & ~$m$~ & ~$m+1/2$~
& ~$m+1/4$~ & ~$m+3/4$ & ~$m$~ & ~$m+1/2$~ & ~$m+1/4$~ & ~$m+3/4$~
\\ \hline
\end{tabular}
\caption{The 10 primary fields in the $\mathbb{Z}_4$ parafermion CFT with their conformal dimensions (also conformal spins) and quantum dimensions. The possible values for $\omega$ in the vertex operator is determined by having a singlevalued OPE between the CFT operators for the anyon and the electron. Here, $m\in\mathbb{Z}$. The corresponding anyon has charge $2\omega/3$ (in the unit of $e$).}
\label{tab:Z4}
\end{center}
\end{table*}

\section{Gapped interface between Halperin-330 and Read-Rezayi states}
\label{sec:anyon-con}

With the above review, we are now ready to discuss the interface between the Halperin-330 state and the RR$_4$ state (the interface is abbreviated as the 330-RR$_4$ interface below). Since both edges of the QH states have the same central charge but with opposite chiralities at the interface, the interface has a zero chiral central charge. This kind of interface is gappable, but it does \textit{not} mean that they must be gapped. As a highly relevant example, the edge of the $\mathbb{Z}_2$ toric code (i.e., the boundary between it and the vacuum) remains gapless if the chosen boundary does not break the translational symmetry~\cite{Kou2013, Tam2022}. However, the edge in a generic situation (with the so-called smooth or rough boundary) is gapped due to the condensation of either one of the self-bosons, $e$ or $m$ on the edge~\cite{Kitaev1998, Kitaev2012}. The discussion suggests that there are two different possible gapped boundaries. When the gapped boundary is obtained by condensing $e$, then $m$ is confined and becomes a boundary excitation, or vice versa. This feature turns out to be relevant in the discussion of fully gapped interfaces between a pair of Moore-Read states~\cite{Teo2020}. Can the 330-RR$_4$ interface be fully gapped? If the answer turns out to be positive, there will be no gapless excitations lying on the interface. Then, it may be possible to transmute the original information carried by certain types of the Abelian anyons completely into topological information stored by anyons in the non-Abelian RR$_4$ liquid. As we will show below, this is indeed possible.

In the following discussion, we first employ the anyon condensation approach to explore the gapfulness of the 330-RR$_4$ interface. We use $\mathcal{A}$ and $\mathcal{B}$ to denote the sets of anyons for the Halperin-330 state and RR$_4$ state, respectively. The possible phases of the interface are determined by the resulting phases from condensing different possible anyons (if it occurs) in $\mathcal{A}\times\bar{\mathcal{B}}$. Here, $\bar{\mathcal{B}}$ indicates the conjugation of $\mathcal{B}$. In the present case, we have
\begin{align}
\mathcal{A}
&=\left\{e^{i\alpha\phi_\uparrow}e^{i\beta\phi_\downarrow}\right\},
\\
\bar{\mathcal{B}}
&=\{\eta e^{i\omega_\eta\phi_l}\}.
\end{align}
Here, $\alpha$, $\beta$ are integers, and $\eta$ denotes the Virasoro primary field in the $\mathbb{Z}_4$ parafermion CFT. For each $\eta$, the possible values of $\omega_\eta$ are given by the last row in Table~\ref{tab:Z4} with $m\in\mathbb{Z}$. Since both electrons in the Halperin-330 state and RR$_4$ state are fermions that have trivial mutual statistics with every anyon in $\mathcal{A}\times\bar{\mathcal{B}}$, $\mathcal{A}\times\bar{\mathcal{B}}$ is a fermionic topological order. Specifically, we introduce the symbol $\Psi_e=e^{3i\phi_\uparrow}$ to denote the electron with pseudospin up for the Halperin-330 state. It is a trivial fermion in $\mathcal{A}\times\bar{\mathcal{B}}$.

\subsection{Determination of Lagrangian subset}

The interface can be fully gapped only if there exists a Lagrangian subset $\mathcal{L}\subset\mathcal{A}\times\bar{\mathcal{B}}$, in which the condensed anyons are the only deconfined anyons in the condensed phase. To identify some possible $\mathcal{L}$, we follow the strategy in Ref.~\cite{Teo2020} by first condensing Abelian anyons. After that, we will analyze the resulting phase and check for the necessity of condensing more anyons in the system to obtain a Lagrangian subset. Finally, we discuss the underlying reason that gives rise to the anyon condensation. It is vital to clarify that one can actually condense non-Abelian anyons, but the strategy below greatly simplifies the analysis. When the Lagrangian subset $\mathcal{L}$ only consists of \textit{Abelian bosons}, then the explicit conditions for having a fully gapped interface are:
\begin{enumerate}[label=(\roman*)]
\item trivial monodromy for all $a\in\mathcal{L}$: $M^{a\bar{a}}_I=1$;
\item trivial mutual statistics for all $a, b\in\mathcal{L}$: $M^{ab}_{a\times b}=1$;
\item confinement for all $a\notin\mathcal{L}$: there exists at least one $b\in\mathcal{L}$ that has a non-trivial braiding phase with $a$.
\end{enumerate}
The condition (i) ensures that $a$ is a boson. If $a$ is its own antiparticle (i.e., $a=\bar{a}$ and $a\times\bar{a}=I$), then $a$ should have an integer conformal spin. When $a$ is not its own antiparticle, then it further requires that $b=a\times a$ also has an integer conformal spin~\cite{Bais-PRB2009, Ellens2014, Burnell-review}. For condition (ii), it implies that the anyon $c=a\times b$ also needs to be a boson, otherwise it needs to be condensed as well~\cite{Ellens2014, Bernevig}. Note that one can also condense fermions (by fusing it with the trivial fermion) in a fermionic topological order~\cite{Chenjie2017}, but this turns out unnecessary in our discussion.

\subsubsection{Separation of charge and neutral sectors}

From Table~\ref{tab:Z4-fusion}, it is clear that the four Abelian anyons in the $\mathbb{Z}_4$ parafermion CFT are $I$, $\psi_1$, $\psi_2$, and $\psi_3$. To deduce the set of condensable Abelian bosons, one can in principle analyze all anyons with $\eta=\left\{I, \psi_1, \psi_2, \psi_3\right\}$ and generic values for $\alpha$, $\beta$, and $\omega_\eta$ in the vertex operators. However, this general treatment turns out to be unphysical. As we will discuss in Sec.~\ref{sec:gap-e-tun}, the gapped interface originates from the gapping or localization of counterpropagating edge modes due to electron and quasiparticle tunneling process. Therefore, the charge modes and neutral modes should obtain expectation values independently. This suggests us to separate the charge and neutral sectors in the problem, which is achieved by introducing a new set of modes for the Halperin-330 edge:
\begin{align}
\phi_r=\phi_\uparrow+\phi_\downarrow,
\\
\phi_n=\phi_\uparrow-\phi_\downarrow.
\end{align}
They correspond to the overall charge mode and neutral spin mode for the Halperin-330 edge, respectively. Both modes are right-moving. Using the new set of modes, the topological term in Eq.~\eqref{eq:L0-330} becomes
\begin{eqnarray} \label{eq:L330-new}
L_0=-\frac{3}{8\pi}\left(\partial_t\phi_r\partial_x\phi_r
+\partial_t\phi_n\partial_x\phi_n\right).
\end{eqnarray} 
The $t$ vector becomes $t=(1,0)^T$, which verifies that $\phi_n$ is a neutral mode. Furthermore, we have
\begin{eqnarray}
\mathcal{A}\times\bar{\mathcal{B}}
=\left\{e^{iQ\phi_r} e^{iS\phi_n}\right\}
\times
\{\eta e^{i\omega_\eta\phi_l}\}.
\end{eqnarray}
Here, $Q=(\alpha+\beta)/2$ and $S=(\alpha-\beta)/2$. Since $\alpha\in\mathbb{Z}$ and $\beta\in\mathbb{Z}$, both $Q$ and $S$ can only take half-integer or integer values. The corresponding anyon in $\mathcal{A}$ has charge $2Q/3$ (in the unit of $e$). Furthermore, the conformal spins for the two vertex operators $\exp{(iQ\phi_r)}$ and 
$\exp{(iS\phi_n)}$ are
\begin{eqnarray}
s_Q=h_Q=\frac{Q^2}{3}
~,~
s_S=h_S=\frac{S^2}{3}.
\end{eqnarray}
Each of them will change by an integer value whenever $Q$ or $S$ is changed by $3\mathbb{Z}$. Physically, this corresponds to the fusion between the anyon with a trivial boson in the Halperin-330 state. Thus, we can identify a pair of anyons with 
$(Q, S)$ and $(Q+3\mathbb{Z}, S+3\mathbb{Z})$. More explicitly, one can rescale the fields $\varphi_r=\phi_r/2$ and $\varphi_n=\phi_n/2$ and rewrite Eq.~\eqref{eq:L330-new} as
\begin{eqnarray} \label{eq:L330-rescale}
L_0=-\frac{6}{4\pi}\left(\partial_t\varphi_r\partial_x\varphi_r
+\partial_t\varphi_n\partial_x\varphi_n\right).
\end{eqnarray} 
The possible vertex operators take the form $\exp{(ip\varphi_r)}$ and 
$\exp{(iq\varphi_n)}$, where both $p$ and $q$ can only take integer values now. Thus, both $\varphi_r$ and $\varphi_n$ (and hence the original fields, $\phi_r$ and $\phi_n$) are compactified bosons in the U(1)$_6$ CFT which has $\exp{(6i\varphi)}$ as the trivial boson. For later discussion, the six primary fields involving the neutral spin mode in the U(1)$_6$ CFT  are summarized in Table~\ref{tab:U_6}. Moreover, a pair of anyons with $(Q,S)$ and $[Q+3(\mathbb{Z}+1/2), S+3(\mathbb{Z}+1/2)]$ differ by the fusion of an odd multiples of electrons in the Halperin-330 state. 

\begin{table} [htb]
\begin{center}
\begin{tabular}{|c|c|c|c|}
\hline
~Symbol~ & ~Vertex operator~ & ~Conformal spin~ 
\\ \hline
~$\mathcal{V}_0$~ & ~$1$~  & ~$0$~ 
\\ \hline
~$\mathcal{V}_1$~ & ~$\exp{(i\phi_n/2)}$~ & ~$1/12$~ 
\\ \hline
~$\mathcal{V}_2$~ & ~$\exp{(i\phi_n)}$~ &  ~$1/3$~ 
\\ \hline
~$\mathcal{V}_3$~ & ~$\exp{(3i\phi_n/2)}$~ & ~$3/4$~ 
\\ \hline
~$\mathcal{V}_4$~ & ~$\exp{(2i\phi_n)}$~ &  ~$1/3$~ 
\\ \hline
~$\mathcal{V}_5$~ & ~$\exp{(5i\phi_n/2)}$~ & ~$1/12$~ 
\\ \hline
\end{tabular}
\caption{The six different primary fields in the U(1)$_6$ CFT for the neutral spin mode $\phi_n$. Here, all vertex operators are normal ordered. Note that two vertex operators are identified if they differ by a bosonic operator. For example, we have $\mathcal{V}_4\sim\mathcal{V}_{-2}$ and $\mathcal{V}_5\sim\mathcal{V}_{-1}$.}
\label{tab:U_6}
\end{center}
\end{table}

\subsubsection{Condensation of Abelian bosons}

Following the above discussion, we look for possible values of $Q$, $\omega_\eta$, and $S$ such that both anyons with operators $\exp{(iQ\phi_r)}\exp{(i\omega_\eta\phi_l)}$ and $\eta \exp{(iS\phi_n)}$ are bosonic. The conformal spin for $\exp{(iQ\phi_r)}\exp{(i\omega_\eta\phi_l)}$ is
\begin{eqnarray}
s(Q,\omega_\eta)=\frac{Q^2-\omega_\eta^2}{3}.
\end{eqnarray}
Therefore, a possible solution for $s(Q,\omega_\eta)=0$ (or alternatively, the null vector for the $K$ matrix describing the charge sector) is $(Q,\omega_\eta)=(1,1)$. Hence, we first condense the following four bosons:
\begin{align}
\label{eq:b0}
b_0
&=e^{im_0(\phi_r+\phi_l)},
\\
\label{eq:b1}
b_1
&=\psi_1 e^{3i\phi_n/2}e^{i(m_1+1/2)(\phi_r+\phi_l)},
\\
\label{eq:b2}
b_2
&=\psi_2 e^{im_2(\phi_r+\phi_l)},
\\
\label{eq:b3}
b_3
&=\psi_3 e^{3i\phi_n/2}e^{i(m_3+1/2)(\phi_r+\phi_l)}.
\end{align}
Here, all $m_i\in\mathbb{Z}$. It is easy to check that all of the above bosons satisfy the conditions (i) and (ii) in the previous discussion. This verifies that all of them can be condensed simultaneously~\cite{Burnell-review}. It is worthwhile for clarifying that being a boson does not mean that it must be condensable. For example, the condensation of non-Abelian bosons can be obstructed by a no-go theorem~\cite{Bernevig-nogo}. Also, a self-dual boson having Frobenius-Schur indicator $\varkappa=-1$ is not condensable~\cite{Ellens2014, Simon-Slingerland}. Nevertheless, none of these limitations applies here.

Since a deconfined anyon must have trivial mutual statistics with all the condensed bosons, we first put $m_2=0$ to eliminate a large set of possibilities. The corresponding monodromy is given by
\begin{eqnarray}
M^{\eta\psi_2}_{\eta\times\psi_2}
=\exp{\left[-(2\pi i)\left(h_{\eta\times\psi_2}-h_\eta-h_{\psi_2}\right)\right]}.
\end{eqnarray}
Hence, all anyons with $\eta=\left\{\sigma_{+}, \sigma_{-}, \chi_{+}, \chi_{-}\right\}$ are confined. This result was used by Barkeshli and Wen in showing that the phase transition from the RR$_4$ state to the Halperin-330 state could be described by the anyon condensation of $\psi_2$~\cite{Wen-PRL2010}. By setting $m_2\neq 0$, the monodromy between $Y=\eta \exp{(iS\phi_n)}\exp{(iQ\phi_r)}\exp{(i\omega_\eta\phi_l)}$ and $b_2$ is given by
\begin{eqnarray}
M^{Yb_2}_{Y\times b_2}
=\exp{\left[\frac{4m_2 \pi i}{3}\left(Q-\omega_\eta\right)\right]}
~,~
m_2\in\mathbb{Z}.
\end{eqnarray}
Here, $\eta=\left\{I, \psi_1, \psi_2, \psi_3, \epsilon, \rho\right\}$. Thus, the set of possible deconfined anyons is further reduced to
$e^{i\omega_\eta(\phi_l+\phi_r)}
e^{3i\gamma\phi_r/2}
e^{iS\phi_n}
\times\left\{I, \psi_1, \psi_2, \psi_3, \epsilon, \rho\right\}$, with $\gamma\in\mathbb{Z}$. This set of anyons have trivial mutual statistics with $b_0$ also. By requiring them to have trivial mutual statistics with $b_1$ and $b_3$, it further reduces the possible set of deconfined anyons to four apparently different classes. For the first class, we have
\begin{align} \label{eq:T1}
\nonumber
\mathcal{T}_1
&=e^{im(\phi_l+\phi_r)}
e^{3i(p+1/2)\phi_r}
e^{i(q+1/2)\phi_n}
\times
\left\{I, \psi_2, \epsilon\right\}
\\ \nonumber
&\sim e^{im(\phi_l+\phi_r)}
\left[e^{3i(\phi_r+\phi_n)/2}\right]
\left\{\mathcal{V}_0, \mathcal{V}_2, \mathcal{V}_4\right\}
\times
\left\{I, \psi_2, \epsilon\right\}
\\  \nonumber
&\sim e^{im(\phi_l+\phi_r)}\Psi_e
\left\{\mathcal{V}_0, \mathcal{V}_2, \mathcal{V}_4\right\}
\times
\left\{I, \psi_2, \epsilon\right\}
\\
&\sim \Psi_e
\left\{\mathcal{V}_0, \mathcal{V}_2, \mathcal{V}_4\right\}
\times
\left\{I, \epsilon\right\}.
\end{align}
All $m$, $p$, and $q$ in the above calculation are integers. It is recalled that 
$\Psi_e=\exp{(3i\phi_\uparrow)}=\exp{[3i(\phi_r+\phi_n)/2]}$ is the electron in the Halperin-330 state. In the second line, we used the fact that $\phi_n$ is a compactified boson in the U(1)$_6$ CFT, and labeled the corresponding vertex operators by the symbols in Table~\ref{tab:U_6}. Moreover, we used the fact that $\exp{(3ip\phi_r)}$ is a trivial boson to make the identification denoted by $\sim$. In the last line, the identification is made by fusing the set of anyons in the third line with the condensed boson $b_2$ with $m_2=-m$. 

Using similar procedures, we determine the second class of deconfined anyons as
\begin{align} \label{eq:T2}
\nonumber
\mathcal{T}_2
&=e^{im(\phi_l+\phi_r)}
e^{3ip\phi_r}
e^{iq\phi_n}
\times
\left\{I, \psi_2, \epsilon\right\}
\\ 
&\sim 
\left\{\mathcal{V}_0, \mathcal{V}_2, \mathcal{V}_4\right\}
\times
\left\{I, \epsilon\right\}.
\end{align}
By setting $\eta=\left\{\psi_1, \psi_3, \rho\right\}$, we have the remaining two classes of deconfined anyons,
\begin{align} \label{eq:T3}
\nonumber
\mathcal{T}_3
&=e^{i(m+1/2)(\phi_l+\phi_r)}
e^{3ip\phi_r}
e^{i(q+1/2)\phi_n}
\times
\left\{\psi_1, \psi_3, \rho\right\}
\\
&\sim e ^{i(\phi_l+\phi_r)/2}
\left\{\mathcal{V}_1, \mathcal{V}_3, \mathcal{V}_5\right\}
\times
\left\{\psi_1, \psi_3, \rho\right\},
\end{align}
and
\begin{align} \label{eq:T4}
\nonumber
\mathcal{T}_4
&=e^{i(m+1/2)(\phi_l+\phi_r)}
e^{3i(p+1/2)\phi_r}
e^{iq\phi_n}
\times
\left\{\psi_1, \psi_3, \rho\right\}
\\ 
&\sim \Psi_e ~e ^{i(\phi_l+\phi_r)/2}
\left\{\mathcal{V}_1, \mathcal{V}_3, \mathcal{V}_5\right\}
\times
\left\{\psi_1, \psi_3, \rho\right\}.
\end{align}
Meanwhile, they are actually equivalent to $\mathcal{T}_2$ and $\mathcal{T}_1$, respectively. This is observed by fusing $\mathcal{T}_3$ and $\mathcal{T}_4$ with either $b_1$ or $b_3$ to obtain
\begin{align}
\mathcal{T}_3 &\sim
\left\{\mathcal{V}_0, \mathcal{V}_2, \mathcal{V}_4\right\}
\times
\left\{I, \epsilon\right\}
\sim\mathcal{T}_2,
\\
\mathcal{T}_4 &\sim 
\Psi_e\left\{\mathcal{V}_0, \mathcal{V}_2, \mathcal{V}_4\right\}
\times
\left\{I, \epsilon\right\}
\sim\mathcal{T}_1.
\end{align}
Therefore, the above four classes of deconfined anyons can be combined into
\begin{align} \label{eq:deconfined-before}
\nonumber
\mathcal{T}
&=\left\{1, \Psi_e\right\} \times
\left\{\mathcal{V}_0, \mathcal{V}_2, \mathcal{V}_4\right\}
\times
\left\{I, \epsilon\right\}
\\ 
&=\left\{1, \Psi_e\right\} \times\mathcal{T}_B.
\end{align}
Since $\mathcal{V}_2$ and $\mathcal{V}_4$ cannot be obtained from any fusion between the bosons in the set $\left\{b_0, b_1, b_2, b_3\right\}$, this set is \textit{not} a Lagrangian subset. In the second line of Eq.~\eqref{eq:deconfined-before}, 
$\mathcal{T}_B$ is a bosonic order with $\mathcal{V}_0 I$ being the trivial boson. The separable form in Eq.~\eqref{eq:deconfined-before} is guaranteed for any \textit{Abelian} fermionic order~\cite{Cano2014, Cheng2019}. The Abelianity of $\mathcal{T}$ becomes transparent after splitting the non-Abelian anyons in $\mathcal{T}_B$. Notice that $(\Psi_e)^2=\exp{(3i\phi_r)}\exp{(3i\phi_n)}\sim 1$. 

\subsubsection{Splitting of non-Abelian anyons and the $\mathbb{Z}_3$ toric code}

Since some of the anyons in the original phase have been identified as the trivial vacuum after the anyon condensation, the unconfined non-Abelian anyons in $\mathcal{T}$ may need to split. Consider the fusion:
\begin{eqnarray}
\epsilon\times\epsilon
=I+ \psi_2+\epsilon
\sim 2I + \epsilon.
\end{eqnarray}
Due to the identification $\psi_2\sim I$, the vacuum appears twice. Consequently,
$\epsilon$ needs to split into two Abelian anyons, $\epsilon=\epsilon_1+\epsilon_2$. This matches the quantum dimension as $2=1+1$. Furthermore, the splitting and the fusion rules are consistent only if
\begin{align}
\label{eq:first-fuse}
&\epsilon_1\times\epsilon_1=\epsilon_2,
\\
&\epsilon_2\times\epsilon_2=\epsilon_1,
\\
&\epsilon_1\times\epsilon_2
=\epsilon_2\times\epsilon_1=I.
\end{align}
Then, the fusion rule $\rho\times\rho=I+\psi_2+\epsilon$ implies that the non-Abelian anyon $\rho$ also needs to split, $\rho=\rho_1+\rho_2$. After the splitting, a possible (but not unique) consistent set of fusion rules are
\begin{align}
&\rho_1\times\rho_1=\epsilon_1~,~
\rho_2\times\rho_2=\epsilon_2~,~
\rho_1\times\rho_2=I,
\\
&\epsilon_1\times\psi_1=\rho_2
~,~
\epsilon_2\times\psi_1=\rho_1,
\\ 
&\rho_1\times\psi_1=\epsilon_2
~,~
\rho_2\times\psi_1=\epsilon_1,
\\ 
&\rho_1\times\epsilon_1=\psi_1,
~,~
\rho_1\times\epsilon_2=\rho_2,
\\ \label{eq:last-fuse}
&\rho_2\times\epsilon_1=\rho_1,
~,~
\rho_2\times\epsilon_2=\psi_3.
\end{align}
Note that $\psi_3\sim\psi_1$ in the condensed phase since they differ by a fusion with $\psi_2$ (i.e, $b_2$ with $m_2=0$). Therefore, the fusion rules involving $\psi_3$ are the same as those involving $\psi_1$. By making the change(s), $\epsilon_1\leftrightarrow\epsilon_2$, or/and $\rho_1\leftrightarrow\rho_2$, one can still obtain a consistent set of fusion rules. This is reasonable as it is impossible to uniquely define or fix the anyons resulting from the splitting. On the other hand, they must be inequivalent. Without loss of generality, we will stick with the set of fusion rules in Eqs.~\eqref{eq:first-fuse} --~\eqref{eq:last-fuse} in the following discussion.

The above discussion shows that $\mathcal{T}_B$ actually consists of nine different \textit{Abelian} anyons, so $\mathcal{T}$ is an Abelian fermionic topological order. Then, what is the topological order $\mathcal{T}$? By labeling the nine anyons in $\mathcal{T}_B$ as shown in Table~\ref{tab:Z3-TC}, it is observed that these anyons take the form of $e^p m^q$, where $p, q\in\mathbb{Z}$. Moreover, one has $e^3=m^3=\mathbb{I}$. Hence,$\mathcal{T}_B$ is the $\mathbb{Z}_3$ toric code. This topological order is a generalization of the more famous $\mathbb{Z}_2$ toric code that only has four anyons, $\left\{1, e, m, f\equiv em\right\}$~\cite{Kitaev2003}. Since it is the topological order built from (mathematically, the modular tensor category constructed over) the quantum double model with the finite group $\mathbb{Z}_3$~\cite{Drinfeld, Kitaev2003, Levin2005}, the $\mathbb{Z}_3$ toric code is also denoted as $\mathfrak{D}(\mathbb{Z}_3)$~\cite{Cong2016, Cong2017, Cong-Math2017, Cong-PRB2017}. It was suggested that this special topological order can be realized by proximitizing a bilayer system of electrons and holes in separate Laughlin states with respective filling factors $\pm 1/3$ to a superconductor~\cite{Barkeshli2016}. It was also pointed out that the more general $\mathbb{Z}_p$ toric code appears as the symmetry-enriched neutral sector of non-diagonal quantum Hall states~\cite{Tam2021}. Here, we have shown that the interface between the Halperin-330 and RR$_4$ states with suitable anyon condensation also leads to the $\mathbb{Z}_3$ toric code. The physical origin that triggers such an anyon condensation will be discussed in Sec.~\ref{sec:gap-e-tun}. On the application side, it was suggested that the $\mathbb{Z}_3$ toric code could be used in implementing universal topological quantum computation~\cite{Cong2017}. To summarize, the condensation of $b_0$, $b_1$, $b_2$, and $b_3$ leads to the condensed phase with deconfined anyons,
\begin{eqnarray}
\mathcal{T}=\left\{1, \Psi_e\right\}\times\mathfrak{D}(\mathbb{Z}_3).
\end{eqnarray}
Note that $\mathcal{T}$ does not describe a fully gapped interface. This is because 
$\mathfrak{D}(\mathbb{Z}_3)$ is not a topologically trivial order. Alternatively, the edge of $\mathfrak{D}(\mathbb{Z}_3)$ is gappable but the edge remains gapless unless a further anyon condensation occurs, which we are going to discuss below.

\begin{table} [htb]
\begin{center}
\renewcommand{\arraystretch}{1.25}
\begin{tabular}{|c|c|c|c|}
\hline
~Symbol~ & ~Anyon in $\mathcal{T}_B$~ & ~Conformal spin $(s=h-\bar{h})$~ 
\\ \hline
~$\mathbb{I}$~ & ~ $\mathcal{V}_0 I$~ & ~$0$~
\\ \hline
~$e$~ & ~$\mathcal{V}_2\epsilon_1$~  & ~$0$~ 
\\ \hline
~$e^2$~ & ~ $\mathcal{V}_4 \epsilon_2$~ & ~$0$~
\\ \hline
~$m$~ & ~$\mathcal{V}_4\epsilon_1$~  & ~$0$~ 
\\ \hline
~$m^2$~ & ~$\mathcal{V}_2\epsilon_2$~  & ~$0$~ 
\\ \hline
~$e^2m$~ & ~$\mathcal{V}_2 I$~  & ~$1/3$~ 
\\ \hline
~$em^2$~ & ~$\mathcal{V}_4 I$~  & ~$1/3$~ 
\\ \hline
~$em$~ & ~$\mathcal{V}_0 \epsilon_2$~  & ~$2/3$~ 
\\ \hline
~$e^2m^2$~ & ~$\mathcal{V}_0 \epsilon_1$~  & ~$2/3$~ 
\\ \hline
\end{tabular}
\caption{The identification between the nine different anyons in the $\mathbb{Z}_3$ toric code (denoted as $\mathfrak{D}(\mathbb{Z}_3)$ in the previous literature and this work) and the corresponding anyons in the bosonic phase $\mathcal{T}_B$. This identification is made such that the topological twist for $e^p m^q$ is 
$\theta=\exp{(2pq\pi i/3)}$. Also notice that we can always define the conformal spin as positive, since it is defined only up to modulo one.}
\label{tab:Z3-TC}
\end{center}
\end{table}

\subsubsection{Two different phases of gapped interfaces}

As we have shown, the set $\left\{b_0, b_1, b_2, b_3\right\}$ is not a Lagrangian subset, and the condensation of this set of bosons does not lead to a fully gapped interface. On the other hand, a fully gapped interface can be achieved by further condensing some bosonic particles in $\mathfrak{D}(\mathbb{Z}_3)$. By recycling the results in Refs.~\cite{Cong2017}, one can immediately conclude that there are two different types of fully gapped interfaces between the Halperin-330 FQH state and the RR$_4$ FQH state.

The first possible kind of fully gapped interface is obtained by a further condensation of $e$ and $e^2$ in $\mathfrak{D}(\mathbb{Z}_3)$. It is straightforward to verify that both of them satisfy conditions (i) and (ii) in the previous discussion. This implies that they can be condensed simultaneously. Moreover, their condensation leads to the confinement of the remaining anyons (except the identity $\mathbb{I}$) in 
$\mathfrak{D}(\mathbb{Z}_3)$. Therefore, we determine the first Lagrangian subgroup~\cite{footnote-group} for $\mathcal{A}\times\bar{\mathcal{B}}$ as
\begin{eqnarray} \label{eq:Le}
\mathcal{L}_e
=\left\{b_0, b_1, b_2, b_3\right\}\times\left\{\mathbb{I}, e, e^2\right\}.
\end{eqnarray}
The corresponding fully gapped interface is usually known as the $e$-boundary. Notice that the symbol $\times$ actually means picking an element from each set on the right hand side of Eq~\eqref{eq:Le} and then fuse them (see Eq.~\eqref{eq:fusex} for example). The product structure of $\mathcal{L}_e$ ensures that it is a maximal set of condensable bosons.

Another possible type of fully gapped interface is obtained by condensing $m$ and $m^2$ in $\mathfrak{D}(\mathbb{Z}_3)$ instead. This leads to the second Lagrangian subgroup,
\begin{eqnarray} \label{eq:Lm}
\mathcal{L}_m
=\left\{b_0, b_1, b_2, b_3\right\}\times\left\{\mathbb{I}, m, m^2\right\}.
\end{eqnarray}
The corresponding fully gapped interface is known as the $m$-boundary. For both $e$- and $m$-boundaries, the remaining deconfined anyons outside the Lagrangian subgroups are $\mathcal{F}_0=\left\{1, \Psi_e\right\}$. This is the trivial fermionic topological order, which indicates that the condensation of anyons in $\mathcal{L}_e$ or $\mathcal{L}_m$ leads to a fully gapped interface~\cite{Chenjie2017}.

\subsection{Gapped interface from the perspective of electron and quasiparticle tunneling}  \label{sec:gap-e-tun}

While anyon condensation has provided a systematic and mathematical approach in studying the gapfulness of the interface, it will be also desirable to understand the gapping of the interface in a more physical picture. We claim that the anyon condensation may originate from the electron and quasiparticle tunneling processes at the interface in two different steps.

At the beginning, the Halperin-330 and Read Rezayi states are two topologically distinct phases. Therefore, only electrons can tunnel across the two different FQH liquids. In general, the electron tunneling process is described by the following Lagrangian density,
\begin{align} \label{eq:e-tun}
\nonumber
L_\text{el, tun}
=&~\xi_{1,a}(x)\left(\psi_1 e^{3i\phi_l/2}e^{3i\phi_\uparrow}\right)^a
\\ \nonumber
&+\xi_{2,a}(x)\left(\psi_1 e^{3i\phi_l/2}e^{3i\phi_\downarrow}\right)^a
+\text{H.c.}
\\
=&~\xi_{1,a}(x)\left[\psi_1 e^{3i\phi_n/2} e^{3i(\phi_l+\phi_r)/2}\right]^a
\\ \nonumber
&+\xi_{2,a}(x)\left[\psi_1 e^{-3i\phi_n/2} e^{3i(\phi_l+\phi_r)/2}\right]^a
+\text{H.c.}
\end{align}
Note that H.c. stands for the Hermitian conjugation. We include the exponent $a>1$ to describe multi-electron tunneling processes. As the electron operators $\psi_1 e^{3i\phi_l/2}$ and $e^{3i\phi_\sigma}$ (where $\sigma=\uparrow, \downarrow$) enter the same number of times, $L_\text{el, tun}$ conserves the total electric charge. Furthermore, the tunneling amplitudes $\xi_{1,a}(x)$ and $\xi_{2,a}(x)$ are random functions in $x$. This is because the electron tunneling generally does not conserve momentum, and disorder needs to be involved. Due to the random nature of the tunneling, some of the processes are actually irrelevant in the renormalization group sense~\cite{footnote-random}. In this situation, we will need to assume the tunneling strength $W_a$, defined as $\overline{\xi_a(x)\xi_a(x')}=W_a\delta(x-x')$, is sufficiently large so that the charge modes $3a[\phi_l(x)+\phi_r(x)]/2$ can still be pinned. Then, this combination of counterpropagating charge modes obtains a nonzero expectation value, and indicates the localization (analogous to mode gapping in nonrandom tunneling) of charge modes. For the neutral sector, it is expceted that the combination $\psi_1 e^{\pm 3i\phi_n/2}$ will also be gapped by $\mathcal{L}_\text{el, tun}$ in the strong coupling regime. By defining $(\psi_1)^2\sim\psi_2$, $(\psi_1)^3\sim\psi_3$, and $(\psi_1)^4\sim I$, the tunneling processes described by $\mathcal{L}_{\text{el, tun}}$ lead to the condensation of $b_0$, $b_1$, $b_2$, and $b_3$ in Eqs.~\eqref{eq:b0} -- \eqref{eq:b3}. 

After the above condensation or charge-mode localization, the interface remains gapless. Meanwhile, quasiparticles with charges $e/3$ and $2e/3$ can tunnel across this ``new" gapless interface. Let us specifically consider the tunneling of charge $e/3$ anyon with the topological sector $\rho$ in the Read-Rezayi liquid. This process can be described by the following Lagrangian density,
\begin{align} \label{eq:qp-tun}
\nonumber
&~L_\text{e/3, tun}
\\ \nonumber
=&~\zeta_{1}(x)\left(\rho e^{i\phi_l/2}e^{i\phi_\uparrow}\right)
+\zeta_{2}(x)\left(\rho e^{i\phi_l/2}e^{i\phi_\downarrow}\right)
+\text{H.c.}
\\
=&~\zeta_{1}(x)\left[\rho \mathcal{V}_1 e^{i(\phi_l+\phi_r)/2}\right]
+\zeta_{2}(x)\left[\rho \mathcal{V}_5 e^{i(\phi_l+\phi_r)/2}\right]
+\text{H.c.}
\end{align}
Since the charge modes have been localized, we focus on the neutral sector. Now, both combinations $\rho\mathcal{V}_1$ and $\rho\mathcal{V}_5$ have zero conformal spins (see Tables~\ref{tab:Z4} and~\ref{tab:U_6} for reference), which indicates that they are bosonic. In principle, they may be condensed or gapped as well. Consider the set of anyons generated from any fusion between $\rho \mathcal{V}_1$ and 
$\rho \mathcal{V}_5$, we have
\begin{eqnarray}
\left\{\rho \mathcal{V}_1, \rho \mathcal{V}_5\right\}
\times
\left\{\rho \mathcal{V}_1, \rho \mathcal{V}_5\right\}
=\left\{I, \psi_2, \epsilon\right\}
\times\left\{\mathcal{V}_0, \mathcal{V}_2, \mathcal{V}_4\right\}
\end{eqnarray}
A further fusion with $\left\{\rho \mathcal{V}_1, \rho \mathcal{V}_5\right\}$ gives
\begin{eqnarray}
\left\{\psi_1, \psi_3, \rho\right\}
\times\left\{\mathcal{V}_1, \mathcal{V}_3, \mathcal{V}_5\right\}.
\end{eqnarray}
Hence, the fusion between $\rho \mathcal{V}_1$ and $\rho \mathcal{V}_5$ generates the list of anyons (more precisely, their neutral sectors) in $\mathcal{L}_e$ and $\mathcal{L}_m$. In this sense, the condensation described by the Lagrangian subgroup 
$\mathcal{L}_e$ or $\mathcal{L}_m$ may be understood as originating from the gapping of modes due to $e/3$ quasiparticle tunneling at the interface. However, this is only possible if we can treat the whole system as a single topological phase. Otherwise, only electrons can tunnel across two topologically distinct phases (for example, between a FQH liquid and a normal metal)~\cite{Wen-book}.

\section{Topological quantum information scrambling via anyon transmutation}
\label{sec:scrambling}

From the previous section, we found that it is possible (at least in principle) to form a gapped interface between the Halperin-330 state and the Read-Rezayi state at level four. Although both states may describe the FQH state in a bilayer system at total filling factor $2/3$, they host different sets of anyons. Specifically, the Halperin-330 state is a two-component state, in which the anyons possess the layer or pseudospin degree of freedom that is absent in the RR$_4$ state. Meanwhile, the RR$_4$ state supports non-Abelian anyons that have more complicated fusion rules than the usual Abelian quasiparticles. This observation leads to a natural question: What happens when an Abelian quasiparticle from the Halperin-330 liquid crosses the gapped interface and enters the non-Abelian RR$_4$ FQH liquid? In the opposite direction, what happens when we drag a non-Abelian quasiparticle from the RR$_4$ liquid to the Halperin-330 liquid? The answers to these questions lead to the ideas of topological quantum information scrambling and Andreev-like reflection of non-Abelian anyons. 

An anyon $a$ originally in the topological phase $\mathcal{A}$ can pass through the interface (described by the condensed phase of $\mathcal{A}\times\bar{\mathcal{B}}$) and transmutes into an anyon $b$ in $\mathcal{B}$ if and only if $ab$ is a deconfined anyon in the condensed phase. This idea was introduced to study the transmutation between anyons in the interface between the Pfaffian and non-Abelian spin-singlet FQH states~\cite{Bais-PRL2009, Grosfeld2009}. Now, we employ the same kind of argument to study the transmutation of Abelian anyons in the Halperin-330 FQH liquid when they cross the interface and enter the RR$_4$ liquid. 

\subsection{Transmutation of Abelian charge $e/3$ anyon}

First, an Abelian charge $e/3$ anyon in the Halperin-330 state can have pseudospin up or pseudospin down. We denote them as $(e/3, \uparrow)$ and $(e/3, \downarrow)$, respectively. Their associated CFT operators are
\begin{align}
(e/3, \uparrow)\equiv\exp{(i\phi_\uparrow)}
=\mathcal{V}_1\exp{(i\phi_r/2)},
\\
(e/3, \downarrow)\equiv\exp{(i\phi_\downarrow)}
=\mathcal{V}_5\exp{(i\phi_r/2)}.
\end{align}
When they cross the fully gapped 330-RR$_4$ interface, their transmutation depends on the type of the gapped interface (i.e., whether it is an $e$- or $m$-boundary). 

Suppose the fully gapped interface is described by the $e$-boundary. Then, the deconfined anyons are those in $\mathcal{L}_e$ in Eq.~\eqref{eq:Le}. In particular, we have the following pair of deconfined anyons:
\begin{align} \label{eq:fusex}
e^2\times b_1
&=~\left[\rho_1 \exp{(i\phi_l/2)}\right] \left[\mathcal{V}_1\exp{(i\phi_r/2)}\right],
\\
e\times b_1
&=~\left[\rho_2 \exp{(i\phi_l/2)}\right] \left[\mathcal{V}_5\exp{(i\phi_r/2)}\right].
\end{align}
These are the only two deconfined anyons that involve $(e/3,\uparrow)$ and $(e/3,\downarrow)$. Since $e^2\times b_1$ and $e\times b_1$ are condensed bosons in 
$\mathcal{L}_e$, they are identified as the vacuum sector in the condensed phase. Based on the interpretation in Ref.~\cite{Grosfeld2009}, $(e/3, \uparrow)$ and $(e/3, \downarrow)$ will transmute into $\rho_1 \exp{(i\phi_l/2)}$ and $\rho_2 \exp{(i\phi_l/2)}$ in the RR$_4$ FQH state, respectively. It is important to clarify that these transmuted anyons are now in $\mathcal{B}$, but not $\bar{\mathcal{B}}$. Since $\rho_1\neq\rho_2$, the resulting Abelian anyons are different.

\begin{figure} [htb]
\centering
\includegraphics[width=3.0in]{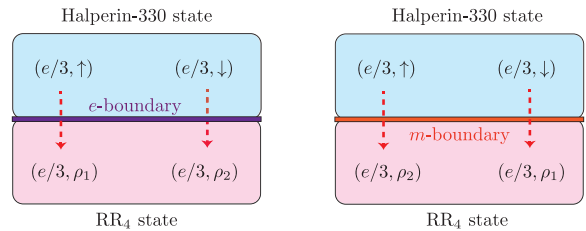}
\caption{The transmutation of an Abelian charge $e/3$ anyon from the Halperin-330 liquid when it crosses the fully gapped interface. The left (right) panel illustrates the case when the interface is described by the $e$-boundary ($m$-boundary). Due to the transmutation, the original local pseudospin information (marked by $\uparrow$ and $\downarrow$) is scrambled, and being stored by the split anyons $\rho_1$ and $\rho_2$. They have different fusion rules with other anyons in the system.}
\label{fig:e3-transmute}
\end{figure}

Consider the case when the fully gapped interface is described by the $m$-boundary that has the Lagrangian subset $\mathcal{L}_m$ in Eq.~\eqref{eq:Lm}. Now, we have the following pair of deconfined anyons:
\begin{align}
m\times b_1
&=~\left[\rho_2 \exp{(i\phi_l/2)}\right] \left[\mathcal{V}_1\exp{(i\phi_r/2)}\right],
\\
m^2\times b_1
&=~\left[\rho_1 \exp{(i\phi_l/2)}\right] \left[\mathcal{V}_5\exp{(i\phi_r/2)}\right].
\end{align}
Thus, $(e/3, \uparrow)$ and $(e/3, \downarrow)$ will transmute respectively into 
$\rho_2 \exp{(i\phi_l/2)}$ and $\rho_1 \exp{(i\phi_l/2)}$. Notice that the result is different from the one in the $e$-boundary. We summarize the above results in Fig.~\ref{fig:e3-transmute}.

\subsection{Transmutation of Abelian charge $2e/3$ anyon}

Similarly, there will be a transmutation of the pseudospin information carried by a charge $2e/3$ anyon when it crosses the interface and enters the RR$_4$ FQH liquid. There are three different pseudospin states for the charge $2e/3$ anyon in the Halperin-330 FQH liquid. They are spin up, spin zero, and spin down. Physically, they can be viewed as the combination of two spin-up, one spin-up and one spin-down, and two spin-down $e/3$ anyons. The corresponding CFT operators are
\begin{align}
(2e/3, \uparrow)
&\equiv\exp{(2i\phi_\uparrow)}
=\mathcal{V}_2\exp{(i\phi_r)},
\\
(2e/3, 0)
&\equiv\exp{[i(\phi_\uparrow+\phi_\downarrow)]}
=\mathcal{V}_0\exp{(i\phi_r)}.
\\
(2e/3, \downarrow)
&\equiv\exp{(2i\phi_\downarrow)}
=\mathcal{V}_4\exp{(i\phi_r)}.
\end{align}

\begin{figure} [htb]
\centering
\includegraphics[width=3.2in]{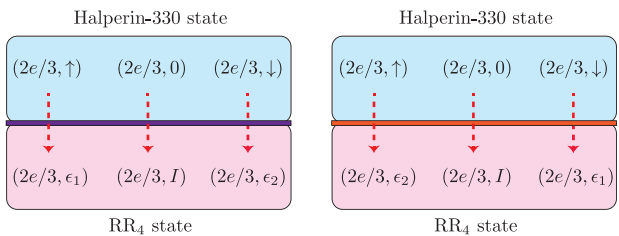}
\caption{The transmutation of an Abelian charge $2e/3$ anyon from the Halperin-330 liquid when it crosses the fully gapped interface. The left (right) panel illustrates the case when the interface is described by the $e$-boundary ($m$-boundary). The original local pseudospin state can be $\uparrow$, $0$, or $\downarrow$. This information is scrambled and being stored by the split anyons $\epsilon_1$, $\epsilon_2$, or the trivial vacuum $I$.}
\label{fig:2e3-transmute}
\end{figure}

Following the previous discussion, we first discuss their transmutation when the fully gapped interface is described by the $e$-boundary. In this case, one has the following three deconfined anyons,
\begin{align}
e\times b_0
&=~\left[\epsilon_1 \exp{(i\phi_l)}\right] \left[\mathcal{V}_2\exp{(i\phi_r)}\right],
\\
\mathbb{I}\times b_0
&=~\left[I \exp{(i\phi_l)}\right] \left[\mathcal{V}_0\exp{(i\phi_r)}\right],
\\
e^2\times b_0
&=~\left[\epsilon_2 \exp{(i\phi_l)}\right] \left[\mathcal{V}_4\exp{(i\phi_r)}\right],
\end{align}
Thus, $(2e/3, \uparrow)$ transmutes into $\epsilon_1 \exp{(i\phi_l)}$; $(2e/3, 0)$ becomes $I \exp{(i\phi_l)}$; $(2e/3, \downarrow)$ becomes $\epsilon_2 \exp{(i\phi_l)}$. For the case of having the $m$-boundary, one has
\begin{align}
m^2\times b_0
&=~\left[\epsilon_2 \exp{(i\phi_l)}\right] \left[\mathcal{V}_2\exp{(i\phi_r)}\right],
\\
m\times b_0
&=~\left[\epsilon_1 \exp{(i\phi_l)}\right] \left[\mathcal{V}_4\exp{(i\phi_r)}\right].
\end{align}
In this scenario, $(2e/3, \uparrow)$ transmutes into $\epsilon_2 \exp{(i\phi_l)}$, whereas $(2e/3, \downarrow)$ becomes $\epsilon_1 \exp{(i\phi_l)}$. The results are illustrated in Fig.~\ref{fig:2e3-transmute}.

\subsection{Topological scrambling of pseudospin information}

The original pseudospin information of a charge $e/3$ or a charge $2e/3$ anyon from the Halperin-330 state is entirely encoded in the neutral spin mode $\phi_n$. Clearly, this is a local information which can be accessed via local measurement. More specifically, we know in which layer the Abelian anyon was created. Depending on \textit{both} the original pseudospin state and the boundary type of the fully gapped interface, the pseudospin information will be transmuted into $\rho_1$ or $\rho_2$ (for the charge $e/3$ anyon), or $\epsilon_1$, $\epsilon_2$, or $I$ (for the charge $2e/3$ anyon). Although one knows that the split anyons, say $\rho_1$ and $\rho_2$, are different, it is impossible to distinguish between them via any local measurement. In fact, their inequivalence is only manifested topologically in their braiding and fusion rules with other anyons (including themselves) as summarized in Eqs.~\eqref{eq:first-fuse} --~\eqref{eq:last-fuse}. In this sense, we claim that the original local pseudospin information is completely scrambled into a form of nonlocal information being stored and protected topologically by the anyons in the RR$_4$ FQH liquid. No information is stored on the interface. It is reasonable as the interface is fully gapped, and does not support any low-energy gapless excitations there. This feature is different from the situation of having a gapless QH interface. As a concrete example, our recent work~\cite{Pf-331} showed that the pseudospin information for an Abelian charge $e/4$ anyon from the Halperin-331 state will be scrambled and stored nonlocally by a pair of vortices (one on the interface, and another one in the Pfaffian FQH liquid) when the Abelian anyon crosses the gapless 331-Pfaffian interface. The comparison here demonstrates the dependence of the scrambling mechanism on the gapfulness of the QH interface.

For an incoming electron from the Halperin-330 liquid, it will simply pass through the interface and become an electron in the RR$_4$ liquid. The latter takes the usual form, $\psi_1 e^{3i\phi_l/2}$ or $\psi_3 e^{3i\phi_l/2}$. The tricky point is that these two are actually identified after the condensation of $\psi_2$ (i.e., $b_2$ with $m_0=0$). Although an electron does not transmute in the present case, other interfaces do allow the transmutation or fractionalization of electrons. For example, the fractionalization of electron at the interface between a $\mathbb{Z}_2$ short-ranged resonating bond quantum spin liquid and a superconductor was proposed~\cite{BBK2014}. In the 331-Pfaffian interface~\cite{Pf-331}, it is feasible for an electron from the Halperin-331 liquid to leave its fermionic nature on the interface by exciting a Majorana fermion there, and becomes a bosonic particle in the Pfaffian liquid.

\subsection{Some further remarks}

One may realize that the transmuted particle from the charge $2e/3$ anyon can be obtained directly by fusing a pair of transmuted particles from the charge $e/3$ anyons. This is not a coincidence. It is guaranteed from the commutativity between restriction and fusion in anyon condensation~\cite{Bais-PRB2009}. This important property also ensures that the total quantum dimension of the split anyons is identical to the quantum dimension of the original non-Abelian anyon. Moreover, the total conformal spin for the transmuted anyon(s) matches the one for the original anyon. In the present case, we have a fully gapepd interface, and the vertex operators describing the charge sector of the deconfined anyons always have zero conformal spins. Thus, the conformal spin for the parafermionic field $\eta$ from the $\mathbb{Z}_4$ CFT must match with the conformal spin for the vertex operator describing the neutral spin mode $\phi_n$. This holds in all our results, and further highlights the advantage of separating the charge and neutral sectors as we did in our analysis.

In principle, one can drag more quasiparticles from the Halperin-330 liquid into the RR$_4$ liquid. Suppose the resulting anyons are well separated from each other (i.e., with a separation much larger than the magnetic length). Then, the braiding between the transmuted anyons will lead to a further scrambling of the original pseudospin information~\cite{Chamon2019}. In order to recover the original pseudospin information, one basically needs to ``pull" all the anyons back to the Halperin-330 liquid. However, the recovered information will be in a highly entangled form, which again resembles the definition of quantum information scrambling.

\section{Andreev-like reflection of non-Abelian anyons}

It is observed that \textit{all} Abelian quasiparticles and quasiholes from the Halperin-330 state can pass through the fully gapped interface, and transmute into a corresponding anyon in the RR$_4$ liquid. However, not all anyons from the RR$_4$ liquid can pass through the interface. This is because the Halperin-330 state can be obtained from the condensation of $\psi_2$ in the RR$_4$ state~\cite{Wen-PRL2010}. From Table~\ref{tab:Z4}, the confined anyons with the parafermionic sector $\eta=\left\{\sigma_+, \sigma_-, \chi_+, \chi_-\right\}$ correspond to quasiparticles in the RR$_4$ FQH liquid that have electronic charges given by odd multiples of $e/6$. This kind of quasiparticles cannot exist in the Halperin-330 state, so they cannot pass through the interface (independent of being gapped or gapless). This further justifies their confinement. On the other hand, a charge $e/6$ anyon may combine with another charge $e/6$ anyon in the RR$_4$ liquid to form a charge $e/3$ anyon, which can then pass through the interface. Another more interesting possibility is that the single non-Abelian $e/6$ anyon strikes the interface, and a charge $-e/6$ anyon with different topological sector is reflected to the RR$_4$ liquid. This also allows the transmission of a charge $e/3$ anyon to the Halperin-330 liquid. This scenario realizes an Andreev-like reflection of non-Abelian anyons in the system, which can be studied systematically from our results in Sec.~\ref{sec:anyon-con}.

\begin{figure} [htb]
\centering
\includegraphics[width=2.0in]{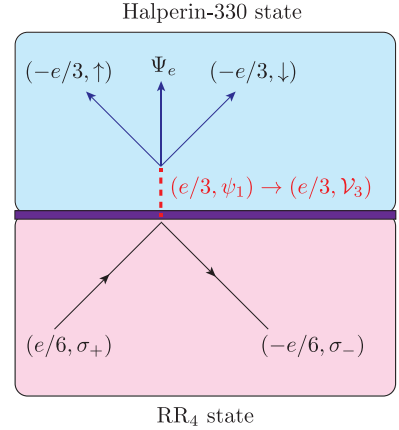}
\caption{A possible Andreev-like reflection of the non-Abelian charge $e/6$ anyon in the fully gapped 330-RR$_4$ interface. The process transmits a charge $e/3$ anyon across the interface by reflecting a charge $-e/6$ anyon back to the RR$_4$ liquid. This is captured by the deconfined boson $b_1$. In the Halperin-330 liquid, the transmitted anyon can further split into an electron, and a pair of charge $-e/3$ anyons with opposite pseudospins.}
\label{fig:Andreev1}
\end{figure}

For example, we consider a charge $e/6$ anyon with its parafermionic sector $\eta=\sigma_+$. Its corresponding CFT operator is $\sigma_+ \exp{(i\phi_l/4)}$. It can be written as
\begin{eqnarray}
\sigma_+ e^{i\phi_l/4}
=\left(\eta_1 e^{i\phi_l/2}\right)\times \left(\eta_2 e^{-i\phi_l/4} \right),
\end{eqnarray}
where $\eta_1\times\eta_2=\sigma_+$ needs to be satisfied. The first term on the right hand side is clearly a charge $e/3$ anyon, and the second term is a charge $-e/6$ anyon. We now explore different possibilities of $\eta_1$ and $\eta_2$, and label the solutions in the form $\left(\eta_1, \eta_2\right)$. From Tables~\ref{tab:Z4-fusion} and~\ref{tab:Z4}, one finds that the possible solutions are 
$(\psi_1, \sigma_-)$, $(\psi_3, \chi_-)$, $(\rho, \sigma_-)$, and $(\rho, \chi_-)$. Then, we can employ our results from anyon condensation to consider the transmutation of the anyon described by $\eta_1 \exp{(i\phi_l/2)}$. For $\eta_1=\psi_1$, we have the deconfined anyon in both $\mathcal{L}_e$ and $\mathcal{L}_m$,
\begin{align} \label{eq:Andreev1}
\nonumber
b_1
&=\psi_1 e^{i\phi_l/2} e^{3i\phi_n/2} e^{i\phi_r/2}
\\ \nonumber
&=\psi_1 e^{i\phi_l/2} \left(e^{3i\phi_n/2} e^{3i\phi_r/2}\right)
\left(\mathcal{V}_5e^{-i\phi_l/2}\right)\left(\mathcal{V}_1 e^{-i\phi_l/2}\right)
\\ \nonumber
&=\psi_1 e^{i\phi_l/2} \left[\Psi_e
\left(e^{-i\phi_\uparrow}\right)\left(e^{-i\phi_\downarrow}\right)\right]
\\ 
&=\psi_1 e^{i\phi_l/2} \left[\Psi_e
\left(-e/3, \uparrow\right)\left(-e/3, \downarrow\right)\right].
\end{align}
Therefore, the corresponding Andreev-like reflection transmits a charge $e/3$ ``composite" anyon in the Halperin-330 liquid and reflects a charge $-e/6$ anyon back to the RR$_4$ liquid. The Abelian $e/3$ composite anyon has an unnatural neutral sector $\mathcal{V}_3=\exp{(3i\phi_n/2)}$, which indicates that it can be further decomposed. From the last line in Eq.~\eqref{eq:Andreev1}, this composite anyon can further ``decay" into an electron $\Psi_e$, and a pair of pseudospin up and pseudospin down $-e/3$ anyons in the Halperin-330 liquid. We illustrate this Andreev-like reflection in Fig.~\ref{fig:Andreev1}. From the deconfined anyon $b_3$ and similar procedures in Eq.~\eqref{eq:Andreev1}, it is straightforward to analyze the case with $\eta_1=\psi_3$ and $\eta_2=\chi_-$. This case is actually identified with the first case since 
$\psi_3=\psi_1\times\psi_2\sim\psi_1$ and $\chi_-=\sigma_-\times\psi_2\sim\sigma_-$ in the condensed phase.  

\begin{figure} [htb]
\centering
\includegraphics[width=2.0in]{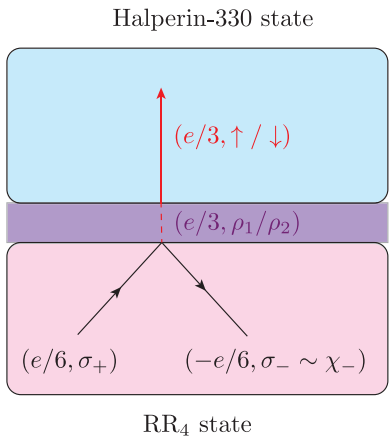}
\caption{Another possible Andreev-like reflection of the non-Abelian charge $e/6$ anyon in the fully gapped 330-RR$_4$ interface. The process creates a charge $-e/6$ anyon being reflected back to the RR$_4$ liquid, and transmits a charge $e/3$ anyon with natural pseudospin state to the Halperin-330 liquid.}
\label{fig:Andreev2}
\end{figure}

For the other two cases $(\rho, \sigma_-)$ and $(\rho, \chi_-)$, it is impossible to distinguish between them and study their associated Andreev-like reflections unambiguously. It is because $\rho$ has already split as $\rho=\rho_1+\rho_2$ in the condensed phase. The two seemingly different fusion rules 
\begin{align}
\rho\times\sigma_-
=(\rho_1+\rho_2)\times\sigma_-
=\sigma_+ +\chi_+,
\\
\rho\times\chi_-
=(\rho_1+\rho_2)\times\chi_-
=\sigma_+ +\chi_+,
\end{align}
are actually identified due to the condensation of $\psi_2$. It is impossible to uniquely determine $\rho_1\times\sigma_-$ and $\rho_2\times\sigma_-$. Nevertheless, the transmitted anyon can only be either $(e/3, \uparrow)$ or $(e/3, \downarrow)$. This is manifested in the deconfined anyons $e\times b_1$ and $e^2\times b_1$ for the $e$-boundary, or $m\times b_1$ and $m^2\times b_1$ for the $m$-boundary. The Andreev-like reflection is illustrated in Fig.~\ref{fig:Andreev2}. Note that we cannot distinguish between the two types of gapped boundaries from the outcome of the Andreev-like reflection. Using similar arguments, one can also explore different Andreev-like reflections for the charge $e/2$ anyon in the interface, which we decide to skip the details here.

\section{Conclusions}

To conclude, we have studied theoretically the interface between the Halperin-330 FQH state and the Read-Rezayi FQH state at level four. Using the technique of anyon condensation, we verify that the interface can be fully gapped. This conclusion is consistent with the expectation from the ``folding trick"~\cite{Kong2014, Hung2015, Chenjie2017}, and the phase transition between the two quantum Hall states triggered by the condensation of $\psi_2$ in the $\mathbb{Z}_4$ parafermion conformal field theory~\cite{Wen-PRL2010, Wen-PRB2012}. A possible physical mechanism for gapping the interface may be arising from electron and quasiparticle tunneling at the interface. Going beyond the gapfulness, the explicitly determined Lagrangian subgroups in Eqs.~\eqref{eq:Le} and~\eqref{eq:Lm} show that there are two different types of fully gapped interface, which has not been elucidated in previous work. They are deeply related to the two different types of gapped boundaries for the $\mathbb{Z}_3$ toric code~\cite{Drinfeld, Kitaev2003, Cong2016, Cong2017, Cong-Math2017, Cong-PRB2017}.

The interface setting further allows us to connect the results from anyon condensation to anyon transmutation and Andreev-like reflection of anyons. In particular, the set of deconfined anyons in the condensed phase (i.e., the phase that describes the fully gapped interface) dictates how an anyon should transmute when it passes through the interface~\cite{Grosfeld2009}. We employed this connection to explore the conversion between the local pseudospin degree of freedom carried by the Abelian anyons in the Halperin-330 state and the resulting anyons split from the non-Abelian anyons in the RR$_4$ state. Specifically, we showed that $(e/3,\uparrow)$ and $(e/3,\downarrow)$ from the Halperin-330 liquid would transmute into $(e/3, \rho_1)$ and $(e/3, \rho_2)$ in the Read-Rezayi liquid. This is possible because $\rho$ splits to two \textit{inequivalent} Abelian anyons $\rho_1$ and $\rho_2$ in the condensed phase. Furthermore, the exact way of transmutation depends on the phase of the fully gapped interface. Since the difference between $\rho_1$ and $\rho_2$ is only manifested in their braiding and fusion rules with other anyons in the system, no local measurement can distinguish between them. A similar anyon transmutation can also occur for the charge $2e/3$ anyons with different pseudospins. Besides being nonlocal, the scrambled pseudospin information is protected topologically. Hence, this kind of anyon transmutation exhibits a topological quantum information scrambling of the originally local pseudospin information. Notice that the scrambling here is different from those discussed in previous work on information scrambling in rational conformal field theory and anyons~\cite{GuQi2016, Chamon2019, Deng2021}. Moreover, our work here shows that no scrambled information is stored on the fully gapped interface. This result is different from the situation in the gapless interface between the Halperin-331 state and Pfaffian state studied in our previous work~\cite{Pf-331}. The comparison highlights the dependence of scrambling mechanism on the gapfulness of the quantum Hall interface. 

Being a confined anyon in the condensed phase, a charge $e/6$ non-Abelian anyon from the RR$_4$ liquid cannot pass through the interface on its own. However, the transmission is allowed by reflecting a charge $-e/6$ anyon with a different topological sector back to the RR$_4$ liquid, and transferring an anyon with an overall charge $e/3$ to the Halperin-330 FQH liquid. Our work shows that this kind of Andreev-like reflection is indeed captured by and can be studied systematically from anyon condensation. Meanwhile, it has been known for a long time that anyons can undergo Andreev-like
reflection at an interface between two FQH states with different filling factors~\cite{Fradkin-PRB1998}. The fully gapped interface between a pair of Laughlin states at different filling factors is a simple example of an Andreev reflector of anyons in quantum Hall interfaces~\cite{Santos2017, Hughes2019}. It was also pointed out in the seminal works~\cite{Safi1995a, Safi1995b, Safi1997, Safi1999a, Safi1999b} that Andreev-like reflection can occur for interacting electrons in one-dimensional wires connected to one-dimensional leads.

Last but not least, let us clarify the relevance and significance of our work. In the parameter range describing realistic GaAs systems, previous numerical work suggested that the Read-Rezayi state may be energetically unfavorable to describe the FQH state in the bilayer system at a total filling factor $\nu=2/3$~\cite{Peterson, Kim}. Meanwhile, graphene-based heterostructures have provided another versatile platform to study different fractional quantum Hall states (for a review, see Ref.~\cite{Dean2020} and the references therein) and the fractional Chern insulator states~\cite{Spanton2018}. Since the microscopic details and energy scales in graphene-based systems are quite different from those in GaAs heterostructures, the nature of FQH states in these two systems can also be different. This may lead to the possibility of realizing both two-component Halperin state and one-component Read-Rezayi state in graphene at the same filling factor through suitable tuning. Furthermore, a recent work~\cite{ZHZ2022} demonstrated numerically that the Halperin-441 state (i.e., the bosonic analogue of the Halperin-330 state) can emerge in the system of two-component hard core bosons at the filling factor $\nu=2/5$. Given the high controllablity and tunability in cold atomic systems, it may be feasible to also realize the Read-Rezayi state, and the corresponding interface between it and the Halperin-441 state there. A main difference is that the corresponding fully gapped interface (if it exists) will be equivalent to the trivial bosonic topological order without a transparent fermionic electron. Finally, Refs.~\cite{Wen-PRB2011, Wen-PRL2010} pointed out that the phase transition between the Halperin-330 and RR$_4$ state is just a particular member of a series of continuous transition between Abelian two-component and non-Abelian one-component quantum Hall states. It will be tempting to generalize our discussion and results to other fully gapped interfaces between those Abelian two-component and non-Abelian one-component quantum Hall states.

\section*{Acknowledgment}

The author would like to thank Professor Kun Yang and Pok Man Tam for their useful comments and suggestions. He also thanks Dr. In\`{e}s Safi for introducing her related works and the useful references. The work by KKWM is supported by the Dirac Postdoctoral Fellowship in the National High Magnetic Field Laboratory, which is funded by the National Science Foundation Cooperative Agreement No. DMR-1644779, and the State of Florida.


\begin{thebibliography}{99}
\bibitem{footnote-spin}
In principle, the spin vector $s$ is another required topological quantum number in characterizing the topological order~\cite{Wen-spin}. We will ignore $s$ and focus on the $K$ matrix and $t$ vector, which is sufficient for the present discussion.
\bibitem{footnote-random}
Since the tunneling process is random, the renormalization group (RG) equation reads 
$dW/d(\ln s)=(3-2\Delta)W$. Here, $s$ is the scaling factor of time and space, and 
$\overline{\xi(x)\xi(x')}=W\delta(x-x')$. For the single electron tunneling process (i.e., $a=1$), the scaling dimension for the operator involving the neutral modes $\psi_1\exp{(i\phi_n)}$ is already $h_n=3/4+3/4=3/2$. The scaling dimension of the operator involving charge modes $\exp{[i(\phi_l+\phi_r)/2]}$ is non-universal but must be positive. Therefore, the process is irrelevant in the RG sense. On the other hand, the corresponding RG equation for nonrandom tunneling is $dW'/d(\ln s)=(2-\Delta)W'$, where $W'$ stands for the tunneling strength. Depending on the interaction between 
$\phi_r(x)$ and $\phi_l(x)$, the single electron tunneling process can be relevant in the RG sense.
\end{thebibliography}
\end{document}